\documentclass[11pt,twoside]{article}
\usepackage{./asp2014}

\aspSuppressVolSlug
\resetcounters

\begin{document}

\title{Be Stars as Seen Through Telescopes in Survey Mode (I)}
\author{Dietrich Baade,$^1$ Christophe Martayan,$^2$ and Thomas Rivinius$^2$
\affil{$^1$European Organisation for Astronomical Research in the Southern Hemisphere, 85748 Garching, Germany; \email{dbaade@eso.org}}
\affil{$^2$European Organisation for Astronomical Research in the Southern Hemisphere, Santiago, Chile; \email{cmartaya@eso.org, triviniu@eso.org}}}

\paperauthor{Dietrich Baade}{dbaade@eso.org}{ORCID_Or_Blank}{ESO}{}{85748 Garching}{}{}{Germany}
\paperauthor{Christophe Martayan}{cmartaya@eso.org}{ORCID_Or_Blank}{}{ESO}{Santiago 19}{}{}{Chile}
\paperauthor{Thomas Rivinius}{trivinius@eso.org}{ORCID_Or_Blank}{ESO}{}{Santiago 19}{}{}{Chile}

\begin{abstract} In spite of the almost all-encompassing variability
of Be stars, surveys play a steadily increasing role in complementing
the insights gained from single-star studies.  The definition of
classical Be stars as recently augmented by \cite{Rivinius2013}
enables unambiguous identification of Be stars in a much increased
range of observations.  Results of targeted surveys are briefly
reviewed for the effects of metallicity, binarity, and evolution.  It
still remains to be seen whether Be stars are safe benchmarks for the
calibration of evolutionary models with rapid rotation.
\end{abstract}

\section{Introduction} 
Single-star studies \cite[e.g.,][]{Stefl2003a,
Stefl2003b, Maintz2003, Stefl2009, Carciofi2009} perhaps still come
close to the maximum of what observations can currently disclose about
(sometimes so-called classical) Be stars.  But the need for
multi-wavelength multi-technique monitoring campaigns renders such
efforts expensive, the necessity of closely contemporaneous
observations can be a major logistical challenge, and the
representativeness of the results can be questionable.

At least as a complementary effort, surveys are very important.
However, the strong irregular variability of key observables makes it
difficult to combine data from different epochs.  Biases abound, and
often surveys are still too exiguous to unveil correlations.  This
situation could be alleviated by efforts to combine different surveys
in metadatabases.  A special statistical pitfall arises from the
unfortunate coincidence that the fraction of Be stars is highest
around spectral type B2, where the Balmer emission is strongest and
the central stars are intrinsically bright.

\section{Surveys}
Be stars are young so that searches should focus on young clusters and
star-forming galaxies.  Less straightforward is the choice of search
criteria.

\subsection{How to identify Be stars?}
For at least one generation, the following definition of Be stars 
\citep{Jaschek1981, Collins1987}
was widely used: ``a non-supergiant star whose spectrum has, or had
at some time, one or more hydrogen / Balmer lines in emission.''  The
strict avoidance of any interpretative risk leaves little leeway but
also leads to confusion with Herbig Be stars, B[e] stars, stars with
magnetospheres, mass exchanging binaries, and even the
meant-to-be-excluded supergiants.

Recently, \citet{Rivinius2013} proposed to add (i) very rapid
rotation, (ii) nonradial pulsation, (iii) the absence of magnetic
fields, and (iv) the irregular ejection (by some unknown process) of
matter, from which a (v) viscous and (vi) dust-free (vii) disk with
(viii) Keplerian rotation is formed.  This opens up eight more
channels for the selection of candidate Be stars, of which, however,
viscosity and the absence of magnetic fields and dust do not seem
suited for positive detections in survey work.  Obviously, this
approach introduces additional options for false positives (also
depending on observing technique).  Examples include:
\begin{list}{$\bullet$}{\partopsep=0mm\itemsep=0mm}
\item
Rapid rotation:  Bn stars
\item
Nonradial pulsation:  SPB stars, surface inhomogeneities, magnetospheres
\item
Ejection of matter (when observed photometrically): gravitational lensing, eclipses
\item
Disk:  magnetospheres, interstellar matter, accretion disks
\end{list}
But since the false positives are mostly rather different from each
other, recursive application of the criteria will lead to the confirmation 
or rejection of the Be-star nature with quickly increasing probability.

In addition to the above, Be stars display characteristic light curves
\citep[e.g.,][]{Mennickent2002}.  While it is unknown whether certain
photometric behaviors are necessarily associated with the Be status of
a star, shapes of light curves can serve as an effective filter.

\subsection{Available surveys and future survey facilities} 
A list will be
submitted to the Web site of the IAU Working Group on Active B
Stars\footnote{\url{http://activebstars.iag.usp.br/}}.

\section{Metallicity effects} 
Metallicity is perhaps the most important parameter targeted by
observations of Be stars beyond the solar neigborhood.  Low
metallicity favors rapid stellar rotation in two ways: (1) Stars are
more compact, and (2) radiative wind driving is less effective so that
stars retain more of their primordial angular momentum.  In agreement
with this expectation, \cite{Martayan2007} and \cite{Martayan2010a}
found that the specific frequency of Be stars and their rotation rates
increase from Milky Way (MW) to LMC and further to SMC.  However, an
only slightly closer look reveals that in all three galaxies peak
cluster-to-cluster variations are not any smaller.  Therefore, there
must be another parameter ranking roughly at par with metallicity.  In
photometry, metallicity effects are often degenerate with the ones of
age.  In fact, Fig.\ 7 in \cite{IqbalKeller2013} is strongly
suggestive of metallicity and age being of comparable relevance for
the abundance of Be stars \citep[see also][]{Maeder1999}.

The interpretation of differences in the frequencies of Be stars in
different environments depends also on the understanding of
metallicity effects on the appearance of their disks, where the
name-giving emission lines form.  On the empirical side
\cite{Wisniewski2007} `speculate' that, at SMC-level metallicity, disks
are either smaller or hotter than in the MW.

Models do not yet come in as firm arbiters.  \cite{HalonenJones2013}
conclude that metallicity-based differences in heating due to
absorption and radiative cooling roughly cancel out.  Be star disks
would only be slightly sensitive to metal abundances.  By contrast,
\cite{AhmedSigut2012} calculate that, at identical stellar effective
temperature, disks with SMC abundances should be hotter by several
thousand Kelvin.  Their models crudely reproduce the distribution of
H$\alpha$ equivalent widths measured by \cite{Martayan2007} in MW and
SMC stars.

\section{Binarity} 
Recent surveys have drastically increased the multiplicity estimates
for massive stars. While \cite{Sana2012} focus exclusively on O stars,
the study by \cite{Chini2012} shows that across the mass range of Be
stars the fraction of multiple stars varies by a factor of a
few. Therefore, it is not obvious whether Be stars as a group are or
are not globally affected.  Snap-shot surveys
\citep[e.g.,][]{Nasseri2013} do not find systematic differences
between Be and other B stars, while short-period systems seem to be
underrepresented among Be stars and there is also no tangible evidence
that passages of a companion trigger mass-loss events.  Theoretical
expectations \citep{Waters1989} that many Be stars should have
white-dwarf companions have not received observational confirmation
\citep{Piters1992}.  But there are three or four Be+sdO systems known
\citep[cf.][]{Peters2013}.  Mass transfer from the progenitor of the
sdO star has probably spun up the Be star and so laid the foundation
to the occurrence of the Be phenomenon \citep[for a description see][]
{Rivinius2013} in these objects.  Because of the low radial-velocity
amplitudes and the difficulty of obtaining UV-to-optical SED's it is
not easy to estimate how statistically important this sub-population
is.  Significant variability in radial velocity of HeII 4686 seems the
best symptom to search for in order to increase the sample.

All in all, the Be phenomenon does have a contribution from
binarity. But the current number and properties of Be stars with a
companion do not indicate that the binary path to today's Be stars is
of elevated importance. This may not have been true, though, at the
time of their formation.

For instance, B stars in high-mass X-ray binaries seem to be fairly
normal Be stars, and they could be the surviving tip of the iceberg
with many others lost in supernova explosions.  However, in that case,
one would expect significantly more high-velocity Be stars, which is
not the case.  Rapid rotation could also be due to early mergers.
Again, observations do not match predictions: Peculiar abundances are
not common in Be stars (Rivinius, Martayan, and Baade, these
proceedings), and of all early-type stars studied, Be stars seem to be
the least magnetic (e.g., Wade, these proceedings).  Therefore, the Be
phenomenon is not a general binary phenomenon.

\section{Evolution of Be stars: models} 
Mastering the challenges of extreme rotation is still an ongoing
process in the development of stellar evolutionary models.  Three
questions play a dominant role: (1) What is different at fast
rotation?  (2) Which predictions can observations verify?  (3)
How can single stars attain, and retain, surface rotation rates as
high as observed in Be stars?

The basic effects are well known \citep{MaederMeynet2012}: The
rotationally reduced equatorial gravity lowers the effective
temperature from pole to equator \citep{vonZeipel1924}.  This causes a
breakdown of the thermal equilibrium, which induces meridional
circulation, and differential rotation adds horizontal turbulence and
shear.  Their combination has a number of consequences:
\begin{list}{$\bullet$}{\partopsep=0mm\itemsep=0mm} 
\item The radial mixing is enhanced.  
\item This increases the effectively available
amount of fuel and so extends the lifetime.  
\item At the same time,
core and envelope are more strongly coupled.  
\item Angular momentum is radially transported and the surface 
rotation accelerated.
\item Surface abundances (especially of helium and nitrogen) are
enhanced.  
\item If and where there is a radiative wind, the mass loss
is strengthened. 
\end{list}

\noindent
Because all Be stars are very fast rotators and readily identified,
models often aim at reproducing the observed properties of Be stars.
But Be stars are special: 
\begin{list}{o}{\partopsep=0mm\itemsep=0mm}
\item Be and Bn stars have different pulsational properties.  
\item Be
stars suffer significant episodic mass loss.  Bn stars do not.  
\item
Be stars show marked star-to-star differences in variability and disk
properties.  
\item Be stars may be particularly little affected by binarity.  
\item Be stars have less relevant large-scale magnetic fields than other early-type stars.  
\end{list} 
That is, Be stars may not be good benchmarks for the calibration of
evolutionary models with fast rotation.  This needs to be kept in mind
throughout the next section, which relies entirely on Be stars.

\section{Evolution of Be stars: observational verification} Perhaps,
the primary result/objective of stellar-evolution models are
evolutionary tracks in the HR diagram.  Because the rotation-induced
differences are mostly relatively subtle \citep{Granada2013},
comparison to observations requires combining data from many clusters,
all of which have different ages, metallicity, foreground reddening,
etc.  Moreover, the transformation from the observational to the
theoretical HRD requires knowledge of the inclination angle of the
rotation axis of each star.  In summary, evolutionary tracks offer
only moderately promising verification opportunities.

A similarly challenging verification channel is the comparison of the
numbers of ultra-rapidly rotating stars.  Models \citep{Granada2013}
predict a clear variation with age - but only in a range, where
reliable measurements can only be done in intensive-care mode for
individual stars.

The evolutionary core-to-envelope angular-momentum transfer process is
slow, and it is fastest in the high-rotation regime
\citep{Granada2013}, where the path to meaningful statistics is long
and laborious.  However, current models simply start in a fully
convective state upon arrival on the ZAMS.  Inclusion of more realistic
star-formation models may lead to more differentiated results.

The clearest finger print of stellar evolution should be imposed on
the surface abundance patterns.  For instance, \cite{Granada2013}
expect rotation to increase the nitrogen surface abundance of a 9
M$_\odot$ SMC star by up to a factor of 500.  Surprisingly,
\cite{Dunstall2011} observed no difference between Be and B stars in
both LMC and SMC.  Accordingly, they concluded that any
critical-rotation phase must be short.  However, this seems at
variance with reports by \cite{Martayan2007} and others that Be stars
already exist as such in the first half of their ZAMS life span.
Taken at face value, this would mean that Be stars are not critical
rotators.  An in-depth investigation of this issue appears critically
needed.

Somewhat surprisingly, observations of late evolutionary stages of
stars with initial masses matching the ones of B stars do not call for
Be stars as progenitors. The only model based on rapid rotation is the
collapsar model by \cite{MacFadyenWoosley1999}, which, however, would
only be applicable to Be stars with an initial mass of more than 15
M$_{\odot}$.  \cite{Martayan2010b} suggested that the number of
high-mass Be stars in the SMC, scaled up to the Universe, could
account for the observed number of gamma-ray bursts.

\section{Initial rotation and its role in the formation of Be stars}
Even if Be stars do not form with rotation rates compatible with the ones
observed later, there are at least four mechanisms that have been proposed to
bridge such a putative gap and lift stars to the high-rotation regime:
\begin{list}{$\bullet$}{\partopsep=0mm\itemsep=0mm}
\item
Merger with a companion
\item
Mass and angular-momentum transfer from a companion
\item
Angular-momentum transport from the shrinking core by meridional
circulation
\item
Angular-momentum transport from the shrinking core by nonradial
pulsation modes \citep{Neiner2013}
\end{list}
However, the single-star cases still require a significant primordial 
reservoir of angular momentum.

In fact, \cite{Martayan2007} find that Be stars arrive on the ZAMS
rotating more rapidly than B stars in general.  That is, their Be
nature may be innate, not acquired.  Furthermore, Be stars are mostly
quasi-single.

This motivates the question whether Be stars can form through a
channel that does not share angular momentum with a companion.  Can
very rapid rotation enable contracting gas clouds to shed excess
angular momentum without fragmentation?  In a way similar to jets in
AGN, microquasars, and collapsars (or to Herbig Haro objects)?

The lower mass limit of Be stars is probably set by the lack of
ionizing photons.  The high-mass limit is usually argued to result
from the onset of strong winds that prevent the formation of disks.
But it is probably necessary to add that strong mass loss also reduces
surface rotation rates, which further quenches any Be attitudes.  On the
other hand, more massive stars are more likely to be multiple.  If Be
stars do not follow this rule, could it, therefore, be that the onset
of fragmentation plays an additional role?

\section{Selected conclusions}
\begin{list}{$\bullet$}{\partopsep=0mm\itemsep=0mm}
\item
Mergers of catalogs would be extremely helpful. Given the 
strong variability on all timescales, the epochs of the observations 
must be included.  This has repercussions on the complexity of the 
database required to support meaningful queries and other operations.  
But the effort would be well spent.  
\item
Degeneracies between metallicity and age require caution in the analysis 
of photometric data.  
\item
As long as it is not confirmed that Be stars are the vanilla type of
rapidly rotating B stars, any investigation of the effects of rapid
rotation should include some rapidly rotating B stars without 
emission lines (Bn stars) for comparison.
\item
The triangle in parameter space between "rapidly rotating", "quasi-single", 
and "magnetic field-free" may be worthwhile exploring for insights 
into the formation of Be stars.  
\end{list}

\noindent This paper is complemented by Rivinius, Martayan, and Baade
(these proceedings).


\begin{thebibliography}{} 
\bibitem[Ahmed and
Sigut(2012)]{AhmedSigut2012} 
Ahmed, A., Sigut, T.A.A.\ 2012, ApJ, 744, 191 
\bibitem[Carciofi et al.(2009)]{Carciofi2009} 
Carciofi, A.C.,Okazaki, A.T., Le Bouquin, J.B.\ et al.\ 2009, A\&A, 504, 915
\bibitem[Chini et al.(2012)]{Chini2012} 
Chini, R., Hoffmeister, V.R., Nasseri, A., Stahl, O., Zinnecker, H.\ 2012, MNRAS, 424, 1925
\bibitem[Collins(1987)]{Collins1987} 
Collins, G.W.\ II 1987, IAU Coll.\ 92, p.\ 3 
\bibitem[Dunstall et al.(2011)]{Dunstall2011}
Dunstall, P.R., Brott, I., Dufton, P.L.\ et al.\ 2011, A\&A, 536, 65
\bibitem[Granada et al.(2013)]{Granada2013} 
Granada, A., Ekstr\"om, S., Georgy, C., et al.\ 2013, A\&A, 553, 25 
\bibitem[Halonen and Jones(2013)]{HalonenJones2013} 
Halonen, R.J., Jones, C.E.\ 2013, ApJS, 208, 3 
\bibitem[Iqbal and Keller(2013)]{IqbalKeller2013} 
Iqbal, S., Keller, S.C.\ 2013, MNRAS, 435, 3103 
\bibitem[Jaschek, Slettebak, and Jaschek(1981)]{Jaschek1981} 
Jaschek, M., Slettebak, A., Jaschek, C.\ 1981, Be Star Newsletter, 4, p.\ 9 
\bibitem[MacFadyen and Woosley(1999)]{MacFadyenWoosley1999}
Mac Fadyen, A.I., Woosley, S.,E.\ 1999, ApJ, 524, 262
\bibitem[Maeder, Grebel, and Mermilliod(1999)]{Maeder1999}
Maeder, A., Grebel, E.K., Mermilliod, J.-C.\ 1999, A\&A, 346, 459
\bibitem[Maeder and Meynet(2012)]{MaederMeynet2012} 
Maeder, A., Meynet, G.\ 2012, Rev.\ Mod.\ Physics, 84, 25 
\bibitem[Maintz et al.(2003)]{Maintz2003}
Maintz, M., Rivinius, Th., \v{S}tefl, S.\ et al.\ 2003, A\&A, 411, 181
\bibitem[Martayan et al.(2007)]{Martayan2007} 
Martayan, Ch., Hubert, A.-M., Floquet, M.\ et al.\ 2007, A\&A, 462, 683 
\bibitem[Martayan, Baade, and Fabregat(2010)]{Martayan2010a} 
Martayan, Ch., Baade, D.,Fabregat, J.\ 2010, A\&A, 509, 11 
\bibitem[Martayan et al.(2010b)]{Martayan2010b}
Martayan, Ch., Zorec, J., Fr\'emat, Y., Ekstr\"om, S.\ 2010, A\&A, 516, 103
\bibitem[Mennickent et al.(2002)]{Mennickent2002} 
Mennickent, R., Pietrzy\'nski, G., Gieren, W., Szewczyk, O.\ 2002, A\&A, 383, 933
\bibitem[Nasseri et al.(2013)]{Nasseri2013} 
Nasseri, A., Dembsky, T., Drass, H., Hoffmeister, V.H., Chini, R.\ 2013, Central European Astroph.\ Bull.,37, 51 
\bibitem[Neiner et al.(2013)]{Neiner2013} 
Neiner, C., Mathis, S., Saio, H., Lee, U.\ 2013, ASPC, Vol.\ 479, p.\ 319 
\bibitem[Peters et al.(2013)]{Peters2013} 
Peters, G.J., Pewett, T.D., Gies, D.R.\ et al.\ 2013, ApJ, 765,2 
\bibitem[Piters et al.(1992)]{Piters1992}
Piters, A.J.M., Meurs, E.J.A., Cote, J.\ et al.\ 2002, IAU Symp. 151, p. 347 
\bibitem[Rivinius, Carciofi, and Martayan(2013)]{Rivinius2013}
Rivinius, Th., Carciofi, A., Martayan, Ch.\ 2013, A\&AR, 21, 69
\bibitem[Sana et al.(2012)]{Sana2012} 
Sana, H.\, de Mink, S.E., de Koter, A.\ et al.\ 2012, Science, 337, 444 
\bibitem[\v{S}tefl et al.(2003a)]{Stefl2003a} 
\v{S}tefl, S., Baade, D., Rivinius, Th.\ et al.\ 2003a, A\&A, 402, 253 
\bibitem[\v{S}tefl et al.(2003b)]{Stefl2003b} 
\v{S}tefl, S., Baade, D., Rivinius, Th.\ et al.\ 2003b, A\&A, 411, 167 
\bibitem[\v{S}tefl et al.(2009)]{Stefl2009}
\v{S}tefl, S., Rivinius, Th., Carciofi, A.C.\ et al.\ 2009, A\&A, 504, 929 
\bibitem[Waters et al.(1989)]{Waters1989} 
Waters, L.B.F.M., Pols, O.R., Hogeveen S.J., Cot\'e, J., v.d.\ Heuvel, E.P.J.\ 1989, A\&A, 220, L1 
\bibitem[Wisniewski et al.(2007)]{Wisniewski2007} 
Wisniewski, J.P.\ 2007, ApJ, 671, 2040 
\bibitem[von Zeipel(1924)]{vonZeipel1924}
von Zeipel, H.\ 1924, MNRAS, 84, 684 
\end{thebibliography}
\end{document}